\documentclass[
reprint,
amsmath,
amssymb,
]{revtex4-2}

\usepackage{graphicx}
\usepackage{dcolumn}
\usepackage{bm}
\usepackage{hyperref}
\usepackage{amsmath}
\usepackage{soul}
\usepackage{xcolor}
\usepackage{hyperref}

\begin{document}
	
	% \title{Memory-induced long-range order in neural activity}
	% \title{Neural activity with memory: long-range order without criticality}
	\title{The Supersymmetric Origin of Chaos and its Hidden Topological Order}\thanks{Published in {\it Newton} (2026), \href{https://www.cell.com/newton/fulltext/S2950-6360(26)00067-8}{doi.org/10.1016/j.newton.2026.100465}}
	
	\author{Igor V. Ovchinnikov}
	 \email{Email: igor.vlad.ovchinnikov@gmail.com}
	 	\affiliation{Device Research Laboratory, Department of Electrical Engineering, University of California at Los Angeles, Los Angeles, 90095, CA, USA}
	 	\affiliation{R\&D, CSD, ThermoFisher Scientific Inc., 200 Oyster Point, South San Francisco, 94080, CA, USA}
	 
	\author{Massimiliano Di Ventra}
	\email{Email: diventra@physics.ucsd.edu}
	
	\affiliation{
		Department of Physics, University of California San Diego, La Jolla, CA, 92093-0319, USA
	}

\begin{abstract}
Dynamical chaos is a term that encompasses a wide range of nonlinear phenomena such as turbulence, neuronal avalanches, weather patterns, and many others. However, despite much work in the field of chaos, its fundamental physical origin still remains not fully understood. In this perspective we report on recent studies showing that chaos is the realization of one of the most fundamental principles in physics: spontaneous symmetry breaking also known as spontaneous ordering. In the present context, the symmetry involved is a {\it topological supersymmetry} inherent to all continuous-time (stochastic) dynamical systems. Chaos is then truly a manifestation of order of topological origin potentially encoding a sort of long-range information hidden beneath its apparent unpredictability. We finally argue that this point of view may have far-reaching implications well beyond chaotic dynamics.
\end{abstract}

	\maketitle

{\bf The scientific concept of chaos is over a century old and can be traced back to the pioneering work of Poincar\'e on celestial mechanics. However, it was only in the 1960s that it began attracting widespread attention, after its rediscovery by Lorenz in numerical experiments on atmospheric dynamics. Since then, it has become clear that chaos is a fundamental phenomenon emerging in virtually all branches of science. This has spawned a flurry of research activities, largely within the understanding that chaos is an unpredictable and apparently \emph{disordered} dynamics. However, recent studies have revealed, quite unexpectedly, that chaos is in fact a manifestation of one of the most foundational concepts in theoretical physics: spontaneous symmetry breaking. In this case, the broken symmetry is a topological supersymmetry that represents the smoothness and continuity of time evolution. Since spontaneous symmetry breaking is synonymous with spontaneous \emph{ordering}, this new realization alters our understanding of chaos at its very core. In this perspective, we present a bird's-eye view of these new developments and explore their potential implications for a broad range of scientific disciplines.}

%%%%%%%%%%%%%%%%%%%%%%%%%%%%%%%%%%%%%%%%%%%%%%%%%%%%%%%%%%%%%%%%%5
%%%%%%%%%%%%%%%%%%%%%%%%%%%%%%%%%%%%%%%%%%%%%%%%%%%%%%%%%%%%%%%%%5
\section{Introduction}
%%%%%%%%%%%%%%%%%%%%%%%%%%%%%%%%%%%%%%%%%%%%%%%%%%%%%%%%%%%%%%%%%5
%%%%%%%%%%%%%%%%%%%%%%%%%%%%%%%%%%%%%%%%%%%%%%%%%%%%%%%%%%%%%%%%%5

What do ferromagnetism, superfluidity, crystallization, and many other physical phenomena have in common? They are all manifestations of one of the most fundamental concepts in physics: spontaneous symmetry breaking (SSB) also known as spontaneous ordering -- a concept central to high-energy and condensed matter physics \cite{Nambu1960,Nambu1961,Goldstone1962}, and beyond (see, e.g., Ref.\cite{RFT_SSB_book}). 

SSB occurs when the ground state of a system has lower symmetry than the underlying equations of motion, giving rise to the emergence of a spontaneous long-range order breaking this symmetry. For instance, in ferromagnets, the spontaneous magnetization breaks rotational symmetry; in superfluidity, the Bose condensate breaks the symmetry associated with particle number conservation; in crystallization, the lattice structure breaks the translational and rotational symmetries of atomic motion; and so on for other instances in which SSB occurs.

\begin{figure}[h]
	\begin{center}
		\noindent\includegraphics[width=1\linewidth]{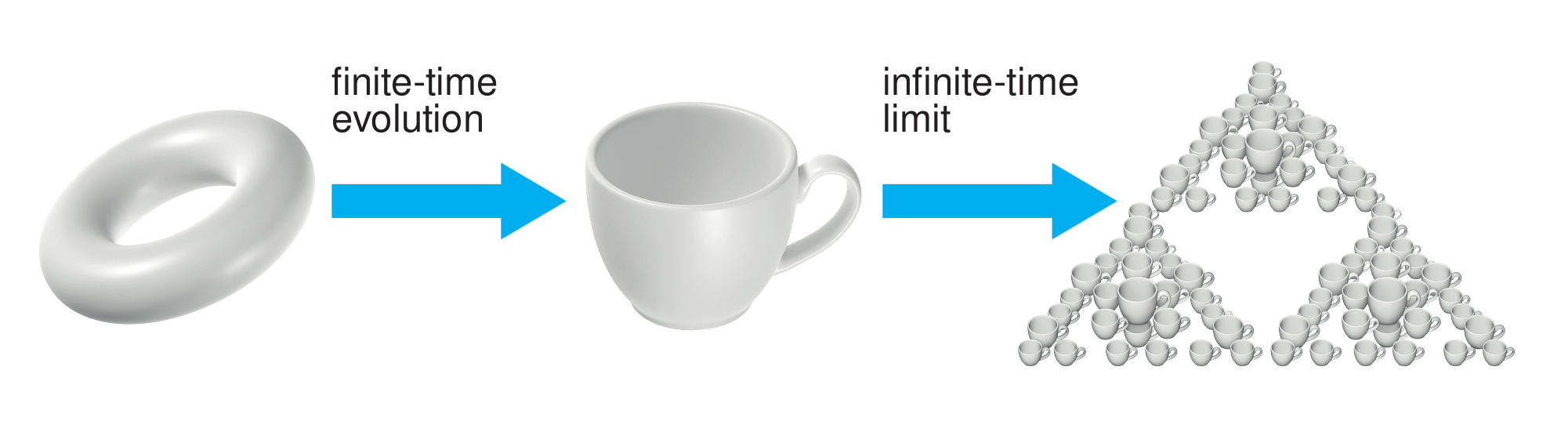}    
	\end{center}
	\caption{\label{Figure_1} {\bf From smooth manifolds to strange attractors} A chaotic or non-integrable  continuous-time dynamics  preserve the topology of smooth manifolds (e.g., in finite time, a torus can smoothly transform into a cup). However, in the limit of infinitely long evolution, it turns these manifolds into fractal objects called ``strange attractors''. The latter ones are not conventional ``shapes'', and they are not topological manifolds from the point of view of algebraic topology. As it is explained in this perspective, the stochastic generalization of this picture corresponds to the spontaneous breakdown of topological supersymmetry.}
\end{figure}

Now, what do chaos and SSB have in common? At face value, the words ``spontaneous order'' and ``chaos'' would not typically be associated with the same phenomenon. In fact, although there is no universal answer to the question of what chaos is, there is general agreement that its main feature is related to the unpredictability of the system's dynamics. This unpredictability simply means that even when initial conditions are chosen arbitrarily close to each other, the corresponding trajectories still diverge in the future. 
In popular parlance, this is often referred to as the ``butterfly effect''.\cite{ButterFly} In addition, chaos is associated with (fractal) structures typically called ``strange attractors'' that are not topological manifolds (see Fig.~\ref{Figure_1}).\cite{Gilmore_book} Both of these features seem difficult to reconcile with the notion of spontaneous order.

Yet, recent research has shown, rather unexpectedly, that chaos is another realization of SSB, hence it is the low-symmetry, {\it ordered} state of the dynamical system showcasing it.\cite{OvcEntropy,Max_2019} In fact, this picture is not too surprising as it reconciles the dual character of the butterfly effect, which can be viewed not only as a mere manifestation of the unpredictability of chaotic dynamics but also as an infinitely long dynamical memory of initial conditions and/or perturbations.\cite{OvcEntropy,Max_2019}

The central question in this picture is then: what symmetry is spontaneously broken in the case of chaotic dynamics? It turns out that such a symmetry is of a fermionic type and of topological nature, namely it is a {\it topological supersymmetry} (TS).
% -- the symmetry pertinent to models called topological field theories.\cite{TFT_BOOK,Witten88,Witten881,Baulieu_1988,Baulieu_1989,labastida1989}

The origin of such a symmetry is even more general than the phenomenon of chaos itself. In fact, it is shared by {\it all} stochastic (partial) differential equations (SDEs). This finding lies at the heart of what can be called supersymmetric theory of stochastics (STS)\cite{OvcEntropy} -- an extension of the supersymmetric approach to stochastic dynamics originally proposed by Parisi and Sourlas.\cite{ParSour1,ParSour} It should be noted, however, that in their work, Parisi and Sourlas did not address the relation between supersymmetry and chaos, as their analysis was restricted to Langevin SDEs, a class of models that are never chaotic. STS generalizes their framework to SDEs of arbitrary form, identifies the spontaneous breakdown of TS as a stochastic generalization of chaos, and yields several further insights, including the clarification that the functional originally proposed by Parisi and Sourlas as the partition function of an SDE is, in fact, a topological object representing the partition function of the noise.
%(Witten index~\cite{Witten88} or Lefschetz number~\cite{Katok-1}) representing 
%topological object known as the Witten index and that was originally proposed by Parisi and Sourlas as the partition function of an SDE is actually a representative of the partition function of the noise.}
%On the mathematical level, it is the topological object known in high-energy physics as the Witten index,~\cite{Witten88} and in dynamical systems theory as the Lefschetz number~\cite{Katok-%1}.}

%As we will explain below, compared to the traditional Fokker-Planck approach to SDEs, STS also contains fermionic variables representing differentials, which are key to consistently account for the butterfly effect. 
Besides being an extension of the Parisi-Sourlas approach, STS can be also viewed as an adaptation of the concept of the generalized transfer operator of the dynamical systems theory \cite{Rue02} to SDEs. 
%like quantum-mechanical systems, also the dynamics of classical systems (whether deterministic or stochastic) can be represented algebraically, namely in terms of a Hilbert space and states in such a space. This means that instead of working directly with the trajectories in the non-linear phase space of a classical system, one can work with state vectors in a linear (Hilbert) space and operators acting on such a space. 
In addition, STS can be recognized as a member of the Witten-type {\it topological field theories} \cite{TFT_BOOK,Witten88,Witten881,Baulieu_1988,Baulieu_1989,labastida1989}, a class of models which feature a TS.
%
%The prize one pays for such an algebraic description of dynamics are that the Hilbert space is infinite-dimensional (to represent a finite non-linear system one always needs an infinite dimensional space) and the evolution operator (the ``Hamiltonian'') is a pseudo-Hermitian 
%operator, namely all its entries are real, and its
%spectrum is composed of only real eigenvalues and pairs of complex
%conjugate eigenvalues.\cite{Mos023} 
%
% Max this price o Hilbert space is paid by the classical Fokker-Plank approach, not STS. STS adds fermions. Lets me rewrite this one a bit
%
%All the peices The price one pays for such an algebraic description of dynamics as compared to the traditional Fokker-Planck approach, is twofold. First, the theory has fermions which represent differentials that define the butterfly effect in numerical experiments. Second, the evolution operator is a pseudo-Hermitian and and its spectrum is composed of only real eigenvalues and pairs of complex conjugate eigenvalues.\cite{Mos023} 
In other words, although different ingredients of STS have been separately proposed across disciplines over the years, it is only very recently that the theory has taken its complete shape and its consequences are starting to emerge.

%Although seeds of such an idea are as old as quantum mechanics, going back to works of Koopman and von Neuman,\cite{Koopman,vonNeumann1} it is only very recently that STS has taken its complete shape and its consequences are starting to emerge.\cite{OvcEntropy}  

%And more recently -- and rather unexpectedly -- SSB has been found to encompass dynamical chaos as well.\cite{OvcEntropy,Max_2019} As pointed out in the context of supersymmetric theory of stochastic dynamics (STS) in Ref. \cite{UTHAMACUMARAN2021100226}: "\emph{...a pioneer of complexity, Prigogine, would define chaos as a spatiotemporally complex form of order...}" and now it turns out to be the right way to look at chaos even from purely mathematical point of view.

%Speaking of chaos, it must first be pointed out that this phenomenon is far more fundamental and pervasive than any other realization of SSB. In physics alone, there are a few concepts that are directly related to it: turbulence, flicker noise, 1/f noise, self-organization etc. And of course chaos appears in all the areas of modern science not only physics. In fact, chaos actually is a mathematical concept and through it, the physical concept of SSB now expands the domain of its relevance from physics to all other scientific disciplines.

In this perspective we bring to the fore this new theoretical development with particular attention to the phenomenon of chaos. We will focus primarily on chaos in continuous dynamical systems since the case of discrete dynamics (maps) is mostly of mathematical interest. We hope to clarify and make it more known the deep connection between chaos and SSB. It must be pointed out from the start, however, that such a development has implications well beyond chaotic dynamics, as it has already found application in astrophysics,\cite{Torsten} neurodynamics,\cite{li2018} and even in new unconventional computing paradigms\cite{di2022memcomputing} where it has revealed topological aspects otherwise difficult to infer from traditional approaches.
%\cite{DMtopo} 
In fact, we are of the opinion that further studies in this direction may bridge the theory of complex systems with other areas of theoretical physics, from quantum field theory to string theory, with possible fruitful cross-fertilization between disciplines that are typically considered non-intersecting.

%is not only interesting but also very promising from the point of view of scientific progress. To highlight its importance from the start, it can be pointed out that the fundamental character of SSB allows to speculate that Richard Feynman would not have called turbulence "the most important \emph{unsolved} problem of classical physics" had he been aware that the hydrodynamic chaos is a realization of SSB.

%%%%%%%%%%%%%%%%%%%%%%%%%%%%%%%%%%%%%%%%%%%%%%%%%%%%%%%%%%%%%%%%%5
%%%%%%%%%%%%%%%%%%%%%%%%%%%%%%%%%%%%%%%%%%%%%%%%%%%%%%%%%%%%%%%%%5
\section{The origin of topological supersymmetry in dynamical systems}%  
%%%%%%%%%%%%%%%%%%%%%%%%%%%%%%%%%%%%%%%%%%%%%%%%%%%%%%%%%%%%%%%%%5
%%%%%%%%%%%%%%%%%%%%%%%%%%%%%%%%%%%%%%%%%%%%%%%%%%%%%%%%%%%%%%%%%5

%Getting more technical, SSB occurs when the ground state of a system has lower symmetry than the underlying equations of motion, due to the emergence of a spontaneous order. For instance, in ferromagnetism, spontaneous magnetization breaks rotational symmetry; in superfluidity, the Bose condensate breaks the global gauge symmetry associated with particle number conservation; and in crystallization, the lattice structure breaks the translational symmetry of atomic motion.

%Going back to chaos, all continuous-time dynamical systems, even in the presence of noise, possess topological supersymmetry (TS) -- algebraic manifestation of the preservation of the proximity of points in the phase space. This is the foundational property of the theory that can be called supersymmetric theory of stochastic dynamics (STS). Within STS, the order associated with the spontaneous breakdown of TS is the dynamical memory of chaotic systems known as the butterfly effect.

%%%%%%%%%%%%%%%%%%%%%%%%%%%%%%%%%%%%%%%%%%%%%%%%%%%%%%%%%%%%%%%%%5
%%%%%%%%%%%%%%%%%%%%%%%%%%%%%%%%%%%%%%%%%%%%%%%%%%%%%%%%%%%%%%%%%5
%\subsection{Deterministic Dynamics and Traditional View on Chaos }
%%%%%%%%%%%%%%%%%%%%%%%%%%%%%%%%%%%%%%%%%%%%%%%%%%%%%%%%%%%%%%%%%5
%%%%%%%%%%%%%%%%%%%%%%%%%%%%%%%%%%%%%%%%%%%%%%%%%%%%%%%%%%%%%%%%%5
\begin{figure}[b]
	\begin{center}
		\noindent\includegraphics[width=1\linewidth]{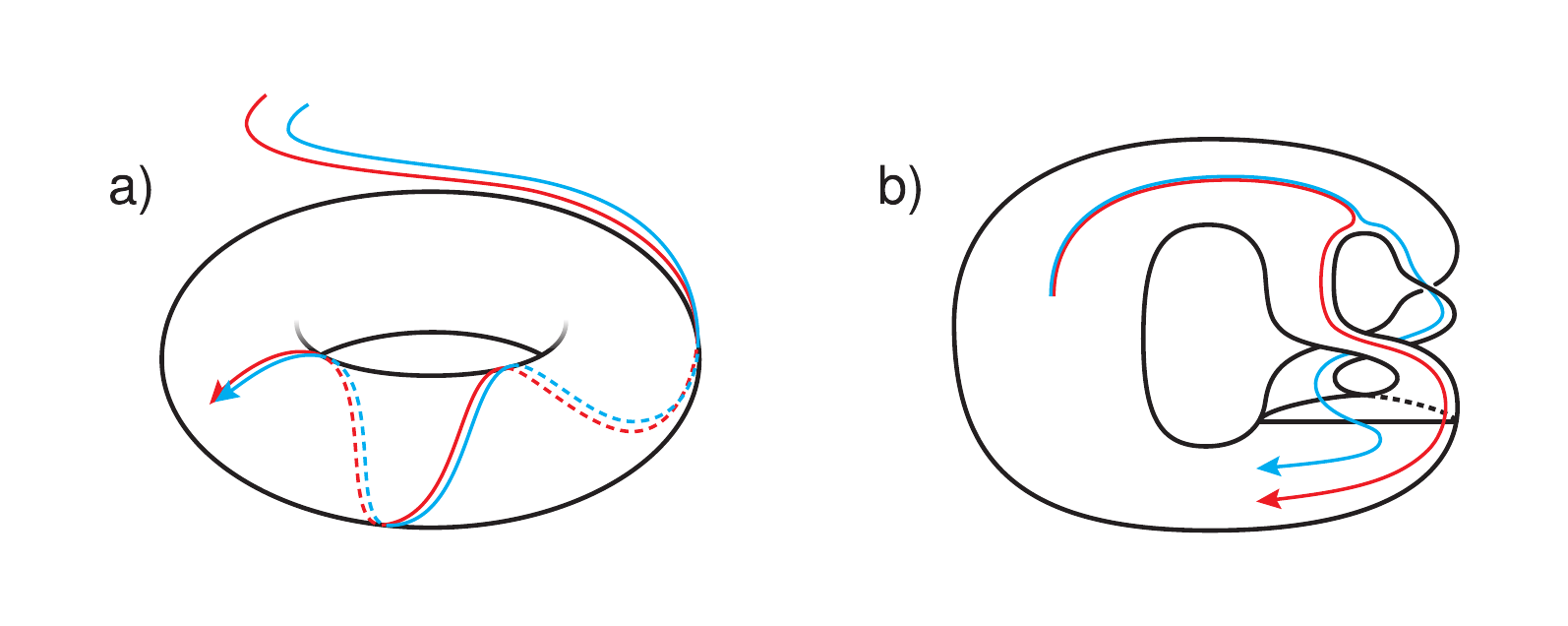}    
	\end{center}
	\caption{\label{Figure_2} {\bf Non-chaotic and chaotic trajectories} 
		{\bf a)} In deterministic dynamics, an integrable (or non-chaotic) system has topologically well-defined attractors such as limit tori and point attractors (the latter ones are not shown). {\bf b)} The ``attractors'' of chaotic (or non-integrable) models are instead fractal objects called ``strange attractors'' (cf. Fig.~\ref{Figure_1}). Their backbone structure is determined by global unstable manifolds,\cite{Gilmore_book} which are branched manifolds with self-intersections. The example given here is the global unstable manifold of a R\"ossler model.\cite{Mangiarotti_Letellier_2021} Contrary to the trajectories of integrable systems that always end up in one of the topologically well-defined attractors and ``forget'' their past -- see {\bf a)}--, close trajectories of chaotic models (blue and red curves) eventually part. This phenomenon, known as the butterfly effect, can be interpreted as infinitely long memory of initial conditions.
	}
\end{figure}

In order to unravel the supersymmetric nature of continuous-time dynamics, let us use deterministic chaotic dynamics as an example. As we have already mentioned, the sensitivity to initial conditions, or butterfly effect, is considered a central property of chaos. It is the ability of the system to separate close initial conditions during time evolution (see Fig.~\ref{Figure_2}). Note that this sensitivity is a ``collective'' property of the trajectories and it cannot be revealed by investigating one trajectory at a time. One must study at least two. For example, the traditional way to reveal it is to investigate two trajectories with close initial conditions. If they eventually part, one can declare that the system exhibits chaotic behavior (Fig.~\ref{Figure_2}).

More rigorously, the butterfly effect can be revealed by defining a differential, $dx$, and propagating it along one trajectory. Namely, if the law of evolution is given by 
\begin{eqnarray}
	%\frac d{dt} 
	\frac{d}{dt} x^i(t) = F^i(x(t)),\label{original}
\end{eqnarray}
where $x^i, i=1,...,D$ are coordinates in a $D$-dimensional phase/state space, with $x=\{x^i\}$, and $F=\{F^i, i=1,...,D\}$ is a flow vector field, then the corresponding evolution of the differential, $dx$, is given by  
%\hl{From now on, summation of repeated indices will be suppressed unless it may lead to a confusion.}
\begin{eqnarray}
	\frac{d}{dt} (dx^i(t)) = TF^i_j(x(t)) dx^j(t),\label{deltas}
\end{eqnarray}
where $TF^i_j(x)=\partial F^i(x)/\partial x^j$, and we have used the summation convention over repeated indices. The finite-time evolution defined by Eq.~(\ref{deltas}) is a matrix whose (logarithms of) eigenvalues are known as Lyapunov exponents.\cite{Gilmore_book} If ``on average'' the largest Lyapunov exponent is positive then close trajectories part in the long-time limit and one concludes that the dynamics are chaotic. If it is negative, then two close trajectories remain close as time goes by. 
%This is actually the preferred method to identify chaos because it is mathematically more rigorous. 

In a similar manner, one can propagate two or more differentials. In doing so, the differentials must be orthogonal to each other because propagating, say, two parallel differentials gives no additional information as compared to propagating only one. From an algebraic point of view, this requirement is fulfilled if the differentials are anticommuting (Grassmann) numbers, or, using physics terminology, fermions, namely $dx^i\sim\chi^i$, with $\chi^i\chi^j=-\chi^j\chi^i$. Products of such differentials represent oriented volumes of different dimensionality,\cite{Nakahara} $\chi^i\chi^j \equiv dx^i \wedge dx^j=-dx^j \wedge dx^i$, and propagating differential volumes provides information on multiple Lyapunov exponents.

We can now introduce the exterior derivative,
\begin{eqnarray}
	\hat d = \sum\nolimits_i \chi^i\partial/\partial x^i,\label{dext}
\end{eqnarray}
the most fundamental operator in algebraic topology.\cite{Nakahara} This operator connects the original equations of motion (\ref{original}) and the equations of motion for the differentials (\ref{deltas}):
\begin{eqnarray}
	%\frac{d}{dt} 
	\frac{d}{dt} x^i(t) = F^i(x(t)) \stackrel{\hat d}{\to} %\frac{d}{dt} 
	\frac{d}{dt} \chi^i(t) = TF^i_j(x(t)) \chi^j(t).
\end{eqnarray}
This observation suggests that $\hat d$ must be a symmetry of any consistent theory that describes temporal evolution of both the original ``bosonic'' variables, $x^i$'s, and their fermionic ``partners'', $\chi^i$'s. It is also clear that $\hat d$ is a \emph{supersymmetry} because it mixes bosonic and fermionic variables. 

%These informal claims will be put onto a more solid mathematical footing in the next section.

%It is worth mentioning for the reader, that the very idea of such a Quantum-Mechanics-like representation of classical dynamics are as old as quantum mechanics itself, going back to works of Koopman and von Neuman.\cite{Koopman,vonNeumann1} But, as can be seen below, STS generalizes these in a few nontrivial ways.

%%%%%%%%%%%%%%%%%%%%%%%%%%%%%%%%%%%%%%%%%%%%%%%%%%%%%%%%%%%%%%%%%%%%%%%%%%%%%%%%%%%5
%%%%%%%%%%%%%%%%%%%%%%%%%%%%%%%%%%%%%%%%%%%%%%%%%%%%%%%%%%%%%%%%%%%%%%%%%%%%%%%%%%%5
\subsection{The Algebraic Description of Continuous-time Dynamics}
%%%%%%%%%%%%%%%%%%%%%%%%%%%%%%%%%%%%%%%%%%%%%%%%%%%%%%%%%%%%%%%%%%%%%%%%%%%%%%%%%%%5
%%%%%%%%%%%%%%%%%%%%%%%%%%%%%%%%%%%%%%%%%%%%%%%%%%%%%%%%%%%%%%%%%%%%%%%%%%%%%%%%%%%5

Of course, unpacking the previous discussion in a mathematically rigorous way requires considerably more work. Since this is a perspective article, we can only give the reader just a flavor of, and the rationale behind these ideas, recognizing that we sweep a lot of details under the rug. To fill in the gaps we must refer the reader to the original papers for an in-depth 
technical account. 

To proceed, let us first realize that deterministic dynamics are just a mathematical idealization. All natural dynamical systems are always subject to the influence of an unpredictable environment. Therefore, stochastic dynamics -- the ones whose equations of motion are subject to external random noise -- represent a more realistic approach to natural systems. In other words, from the mathematical point of view, realistic dynamics are always stochastic. The latter can be described by the following general class of models,
\begin{eqnarray}
	%\frac{d}{dt} 
	\frac{d}{dt} x(t) = F(x(t)) + (2\Theta)^{1/2}\sum\nolimits_a e_a(x(t))\xi^a(t), \label{SDE}
\end{eqnarray}
where $x\in X$ represents the state of the system in the phase/state space, $X$, which is assumed, for simplicity, to be a compact smooth manifold of some dimension $D$, hence in any coordinate neighborhood $x = \{x^i\}$, with $i=1,\dots,D$; $F$ is the flow vector field representing the deterministic law of evolution; $\xi=\{\xi^a, a=1,...D'\}, D'\ge D$ is a set of Gaussian white noises --a convenient, but not necessary choice of noise from the point of view of STS~\cite{Gozzi5}-- each coupled to the model system via a vector field, $e$, and $\Theta$ quantifies the overall strength of the noise so that the deterministic limit corresponds to $\Theta \rightarrow 0$.

However, switching to a stochastic picture of dynamics immediately brings with it one complication. Now, all the trajectories are, in principle, available to the system, much like in quantum dynamics.\cite{Book_Peskin} This means that a description of the dynamics in terms of mere trajectories is insufficient. This is the reason why in traditional approaches to stochastic dynamics, one describes the model in terms of a probability distribution, $P(x)$, that evolves in time according to a Fokker-Planck equation\cite{Risken1996} -- the stochastic analogue of the Sch\"odinger equation. \cite{vanKampen}

This traditional picture of stochastic dynamics can be called a statistical approach because $P(x)$ alone determines the state of the system. It is foundational in Statistical Physics and related areas. It does account for the random character of trajectories. However, the probability distribution does not provide any information on Lyapunov exponents and the corresponding dynamical memory, hence it cannot help us unravel chaotic dynamics. Some generalization of the statistical approach is then needed if we ever want to identify the butterfly effect. 

The essence of this generalization is precisely the introduction of the fermions we have discussed in the previous section in the context of trajectories. More rigorously, this generalization can be done by following the Parisi-Sourlas stochastic quantization method where fermions, called Faddeev-Popov ghosts in quantum field theory, appear as a formal tool coming from what can be interpreted as ``gauge-fixing'' of trajectories to only solutions of SDEs.\cite{TFT_BOOK}  
%to fix fermionic fields appear as formal   
%This generalization is fairly straightforward and borrows a technique typically used in quantum field theory to unambiguously quantize path integrals in the presence of gauge symmetries.\cite{Book_Peskin} We then introduce $D$ additional fields, $\chi=\{\chi^i\}$, (in quantum field theory these are called ``ghost fields'') from the tangent bundle of the phase space, $\chi\in TX$, and instead of working with the probability distribution, $P(x)$, we work with an ``amplitude'' or ``wavefunction'' $\psi(x,\chi)$.  
%At this point, we need to answer an important question. Which statistics of the fields $\chi$'s should we choose? There are only two major choices:  bosonic (or symmetric) and fermionic (or antisymmetric) statistics, in which case $\chi$'s are the so-called Grassmann numbers.\cite{Book_Peskin}  
%A hint to which choice we should make comes from the fact that if we propagate two parallel differentials in time no additional information about Lyapunov exponents (how two nearby trajectories diverge in time) can be obtained. Orthogonality of the differentials then suggests that we should pick fermionic statistics for the fields $\chi$'s. 
However, we will employ a more fundamental approach based on the ideas of dynamical systems theory. It helps unveiling some of the key concepts and it is free of the mathematical ambiguity of the path-integral formulation, known in the context of stochastic dynamics as Ito-Stratonovich dilemma.\cite{vanKampen,Ito_Stratonovich_2025}

We have already provided arguments that in order to track Lyapunov exponents one needs to consider not only the original variables, $x^i$, but also their fermionic counterparts, $\chi^i$. This means that the object that represents the state of the system at a given moment of time -- we may call it the ``wavefunction'' --  must depend not only on $x$, as the probability distribution, $P(x)$, but also on these fermions: 
\begin{eqnarray}
	|\psi(t)\rangle \equiv \psi(t;x,\chi) = \sum\nolimits_{k=0}^{D} (1/k!)\psi^{(k)}_{i_1....i_k}(t;x)\chi^{i_1}...\chi^{i_k},
	\label{psi}
\end{eqnarray}
where the tensors $\psi^{(k)}$'s are antisymmetric in their indices. The right hand side of Eq.~(\ref{psi}) can be looked upon as a Taylor expansion of the ``wavefunction'' in $\chi$ and each term of this expansion is a differential form -- a fundamental concept in algebraic topology.\cite{Nakahara} 

%{\bf This is probably too much mathematical information for a perspective.} \hl{{\bf Not sure we need this but it feels like "engaging the reader"}: Differential forms are naturally coupled or "dual" to submanifolds in $X$: a differential form of degree $k$, $\psi^{(k)}$, can be integrated over a submanifold of dimensionality $k$, $\gamma_k$: $\int_{\gamma_k}\psi^{(k)} \in \mathbb{R}$. Moreover, there is also the concept of Poincare dual, providing a unique differential form for any submanifold. Unlike submanifolds, however, differential forms are linear objects and, particularly, they can be averaged over noise and/or trajectories. It is through this property that STS offers stochastic generalization to some set-theoretic concepts from the dynamical systems theory such as (un)stable manifolds, strange attractors etc.}

%They are naturally coupled to submanifolds of $X$ of corresponding dimensionalities. Differential forms are coordinate free objects as they carry their differentials with them. In this setting, the total probability distribution can be looked upon as the differential form of maximal degree: $\psi^{(D)} = P dx^1  dx^D$, so that the probability of the system to be inside $D$-dimensional volume, $V$, is $\int_V \psi^{(D)}$. Differentials forms of lesser degrees can be looked upon as a generalization of probability distribution.

The evolution of differential forms is governed by what is known in the theory of dynamical systems as ``generalized transfer operator'' (GTO), \cite{Rue02} i.e., transformations of wavefunctions caused by the evolution defined by the SDE and averaged over noise. It can be shown~\cite{OvcEntropy} that
%{\bf Should we take this out? It is probably too technical for a perspective: In our present case, the GTO can be constructed in three steps. First, Eq.~(\ref{SDE}) defines \cite{Slavik} a two-parameter family of noise-configuration-dependent diffeomorphisms, $\hat M(\xi)_{tt'}$, such that the trajectory $x(t)$, starting from $x'$ at time $t'$ is given by $x(t) = \hat M(\xi)_{tt'}(x')$, for each noise configuration $\xi$. Second, the diffeomorphisms induce actions (also known as ``pullbacks'') on the differential forms, $\hat M(\xi)^*_{tt'}$, which define their deterministic noise-configuration dependent evolution $|\psi(t)\rangle = \hat M(\xi)^*_{t't} |\psi(t')\rangle$. 
	%Third, the pullback must be averaged over all the noise configurations. This can be done even if $X$ is nonlinear because the pullback is a linear ``object'' from the mathematical point of view, and the operations of summation of pullbacks and multiplication of a pullback by a number are well defined.} 
%The result of this averaging is the following ``algebraic representation'' of stochastic evolution,
\begin{eqnarray}
	|\psi(t)\rangle = \hat {\mathcal M}_{tt'} |\psi(t')\rangle, 
\end{eqnarray}
with$\;\hat {\mathcal M}_{tt'}$ being the GTO, which takes a particularly simple form for Gaussian white noise, 
\begin{equation}
	\hat {\mathcal M}_{tt'} = e^{-(t-t')\hat H}, \;\quad\; \hat H = \hat L_F + \Theta \sum\nolimits_a \hat L_{e_a} \hat L_{e_a},
	\label{GTO}
\end{equation}
where $\hat L$ are the so-called Lie derivatives along the vector fields specified in the subscript. The operator $\hat H$ above is unique and it equals \cite{Ito_Stratonovich_2025} the stochastic evolution operator in the Stratonovich interpretation of the traditional theory of SDEs.\cite{Elworthy1998,Kunita2019} 
%Note, however, that unlike evolution operators in traditional theory of stochastic dynamics, the above GTO is unique which is a consequence of its most natural definition from the mathematical point of view. This uniqueness can be said to resolve the so-called Ito-Stratonovich dilemma.\cite{Kampen,West}

It is clear from Eq.~(\ref{GTO}) that $\hat H$ plays the role of the generalized Fokker-Planck operator as the ``wavefunction'' evolves according to:
\begin{eqnarray}
	\partial_t |\psi(t)\rangle = -\hat H |\psi(t)\rangle.
\end{eqnarray}
Up to ``imaginary time'', $\hat H$ is similar to the Hamiltonian in Quantum Theory. The difference, however, is that this operator is pseudo-Hermitian.\cite{Mos023} This means that, unlike in Quantum Theory, where the eigenvalues of the Hamiltonian operator are all real, in the present case the eigenvalues of $\hat H$ come in complex conjugate pairs (in the theory of dynamical systems these are called Reulle-Pollicott resonances \cite{Pollicott1985,Ruelle1986}). 
%However, just like in the Hermitian case, except for a few zero-eigenvalue singlets, all eigenstates are doubly degenerate 

%where the \emph{infinitesimal} GTO is the Stratonovich stochastic evolution operator (SEO) in terms of traditional theory of SDEs.\cite{Elworthy1998} Note, however, that due to its most natural mathematical meaning, the GTO is unique and one needs to specify no interpretation of SDE in Eq.(\ref{SDE}).

%%%%%%%%%%%%%%%%%%%%%%%%%%%%%%%%%%%%%%%%%%%%%%%%%%%%%%%%%%%%%%%%%5
%%%%%%%%%%%%%%%%%%%%%%%%%%%%%%%%%%%%%%%%%%%%%%%%%%%%%%%%%%%%%%%%%5
\subsection{Topological Supersymmetry}
%%%%%%%%%%%%%%%%%%%%%%%%%%%%%%%%%%%%%%%%%%%%%%%%%%%%%%%%%%%%%%%%%5
%%%%%%%%%%%%%%%%%%%%%%%%%%%%%%%%%%%%%%%%%%%%%%%%%%%%%%%%%%%%%%%%%5
%So far, we have shown that classical dynamics can be described algebraically, very similarly to quantum mechanics. 
%However, we have yet to identify the topological supersymmetry present in all (stochastic) differential equations. 
Let us now turn to the supersymmetric aspect of stochastic dynamics. In terms of the evolution of the phase/state space, for any configuration of the noise, and smooth enough $F$ and $e$'s in Eq.~(\ref{SDE}), the dynamics defined by an SDE can twist, squeeze, and stretch the fabric of the phase/state space, but they cannot rip it (cf. Fig.~\ref{Figure_1}). In other words, smooth dynamics preserve the proximity of points in $X$, also called topology of $X$: two initially close points will remain close. Conversely, when TS is broken spontaneously, two initial points of the phase/state space may part in the long-time evolution limit represented by a non-supersymmetric ground state. In this way, the spontaneous supersymmetry breaking relates to the traditional understanding of the ``butterfly effect'' (see Fig.~\ref{Figure_2}b and also section~\ref{SecGeneratingFunctional}).

From the algebraic point of view, the presence of supersymmetry in all SDEs follows from the ``naturality of the exterior derivative'',\cite{bott1982differential} namely the fact that the transformations of the differential forms defined by smooth dynamics commute with $\hat d$ in Eq.~(\ref{dext}). Since the GTO is just the corresponding transformation averaged over noise, this means that 
\begin{eqnarray}
	[\hat d, \hat {\mathcal M}_{tt'} ] = 0,
\end{eqnarray}
and by extension from Eq.~(\ref{GTO})
\begin{equation}
	[\hat d, \hat H ] = 0,\label{Hdcomm}
\end{equation}
where we have introduced $[\cdot,\cdot]$ for the so-called bi-graded commutator of two operators.\cite{Nakahara} 
%We also note that in coordinates, the exterior derivative can be written as
%\begin{equation}\label{dext}
	%\hat d = \sum_i \chi^i \frac{\partial}{\partial x^i},
	%\end{equation}
%which shows that $\hat d$ ``creates'' a fermion ($\chi^i$), and ``destroys'' a boson ($x^i$). %In addition, the exterior derivative has the fundamental property of being nilpotent, namely $\hat d^2=0$.

Because of the commutativity of the exterior derivative with $\hat H$, Eq.~(\ref{Hdcomm}), the former can be recognized as a symmetry, or rather a supersymmetry of the model because $\hat d$ ``creates'' a fermion ($\chi^i$), and ``destroys'' a boson ($x^i$) as can be seen from Eq.~(\ref{dext}).
%Accordingly, (most) eigenstates come in doublets, i.e., pairs of the same eigenvalue: if $e^{-(t-t')H_i} |i\rangle = \hat {\mathcal M}_{tt'} |i\rangle$, then $|i'\rangle = \hat d |i\rangle$ is also an eigenstate with the same eigevalue. By standard lore of quantum theory, $\hat d$ can be recognized as a symmetry of the model\hl{When evolution is defined by a linear operator, a presence of symmetry reflects itself as degeneracy of eigenstates} or, rather, a supersymmetry because of its nilpotency, $\hat d^2=0$.

In fact, $\hat d$ is recognized as a {\it topological supersymmetry} (TS). This is because the exterior derivative is the most fundamental operator in algebraic topology -- it is essentially the algebraic version of the boundary operator.\cite{Nakahara} More rigorously, one can show that the algebraic representation of stochastic dynamics we have just discussed is a Witten-type topological field theory (TFT).\cite{Baulieu_1989, Witten88, Witten881, labastida1989} This is because the evolution operator, $\hat H$, can also be written as $\hat H = [\hat d, \hat{\bar d}]$, with $\hat{\bar d} = \hat \imath_F + \Theta \sum_a\hat \imath_{e_a} \hat L_{e_a}$, where $\hat \imath_F$ and $\imath_{e_a}$ are the so-called interior multiplication operators.~\cite{OvcEntropy} This has a similar form as the evolution operator in supersymmetric nonlinear sigma models,\cite{Witten_1982} with the difference that the operator $\hat{\bar d}$ is not necessarily nilpotent. Such form of evolution operator is a definitive property of TFTs.

Like any other symmetry, the presence of a TS leads to degeneracy of eigenstates of $\hat H$. Almost all eigenstates are non-supersymmetric ``doubles'' of the form $|i\rangle$ and $\hat d|i\rangle$. However, there are also a few supersymmetric ``singlets'', $|\theta\rangle$ such that $\hat d|\theta\rangle=0$, and no state $|\theta'\rangle$ exists such that $|\theta\rangle=\hat d |\theta'\rangle$. In mathematical terms, these eigenstates are nontrivial in the so-called de Rham cohomology,~\cite{Nakahara} and there is one state for each de Rham cohomology class because the eigensystem of $\hat H$ is complete. It is also easy to check that the expectation value of any $\hat d$-exact operator, that is of the form $[\hat d, \cdot]$, vanishes on supersymmetric singlets. $\hat H$ is such a $\hat d$-exact operator, so all supersymmetric singlets of $\hat H$ have exactly zero eigenvalue.  

%The way toward TFT is to notice that SEO is a $\hat d$ exact operator:
%\begin{eqnarray}
	%    \hat H = [\hat d, \hat{\bar d}],
	%\end{eqnarray}
%where $\hat{\bar d} = \hat \imath_F + \Theta \hat \imath_{e_a} \hat L_{e_a}$, $\hat \imath_F$ is interior multiplication operator. The above formula follows from Cartan formula saying that Lie derivative is $\hat L_F = [\hat d, \hat \imath_F]$.

%The above form of the evolution operator is similar to that of supersymmetric nonlinear sigma model,\cite{Witten_1982} with the difference that operator $\hat{\bar d}$ is not nilpotent. While there is a way to construct a nilpotent supercharge $\hat d^\ddagger$ such that $\hat H=[\hat d, \hat d^\ddagger]$, the real key difference is pseudo-Hermiticity of $\hat H$. This means that, unlike in quantum theory where eigenvalues are real, the eigenvalues come in complex conjugate pairs. Just like in Hermitian case, however, except of a few zero-eigenvalue singlets, all eigenstates are doubly degenerate.

%%%%%%%%%%%%%%%%%%%%%%%%%%%%%%%%%%%%%%%%%%%%%%%%%%%%%%%%%%%%%%%%%5
%%%%%%%%%%%%%%%%%%%%%%%%%%%%%%%%%%%%%%%%%%%%%%%%%%%%%%%%%%%%%%%%%5
\section{Chaos and Topological Supersymmetry Breaking}
\label{SecGeneratingFunctional}
%%%%%%%%%%%%%%%%%%%%%%%%%%%%%%%%%%%%%%%%%%%%%%%%%%%%%%%%%%%%%%%%%5
%%%%%%%%%%%%%%%%%%%%%%%%%%%%%%%%%%%%%%%%%%%%%%%%%%%%%%%%%%%%%%%%%5
\begin{figure}[t]
	\begin{center}
		\noindent\includegraphics[width=1\linewidth]{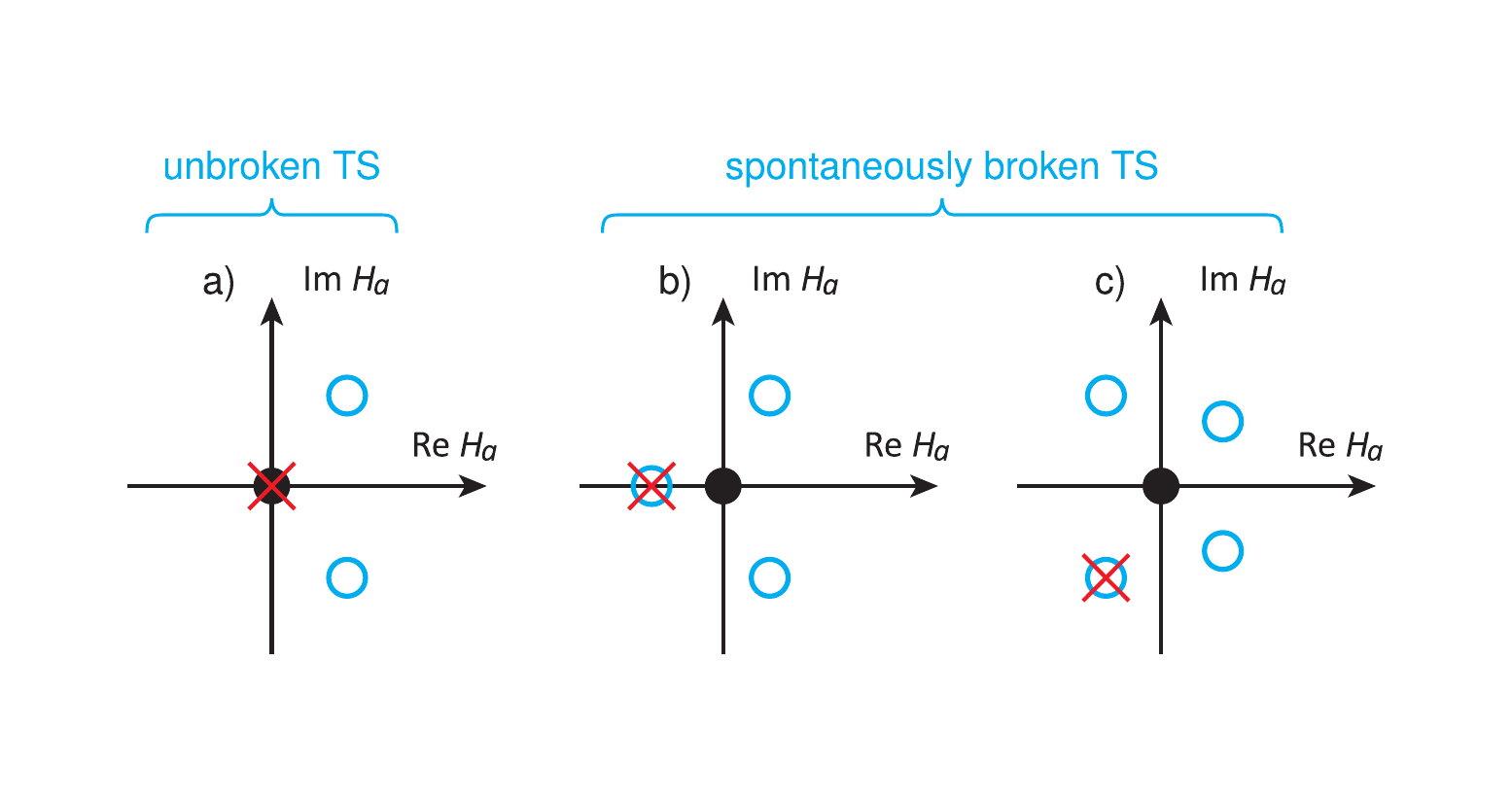}    
	\end{center}
	\caption{\label{Figure_3} {\bf Spectra of the evolution operator} In stochastic dynamics, the notions relevant to chaos are supersymmetric eigenstates (solid black dots at the origin) and the ground states (red crosses) of the evolution operator, $\hat H$, whose possible spectra $\{H_\alpha\}$, with $\alpha$ running over the eigenvalues, are presented in the figure. Supersymmetric eigenstates always exist (for models with compact phase space) and their eigenvalues are exactly zero. The steady-state probability distribution, sometimes called ``ergodic zero'', is one of them. The ground state, in turn, is the stochastic generalization of the concept of attractors. It represents ``sustained dynamics'', i.e., the state of the system after long unperturbed evolution. The ground state is (one of) the eigenstate(s) with the least Re $H_\alpha<0$. For {\bf a)}, the ground state is a supersymmetric eigenstate -- the situation known as unbroken topological supersymmetry (TS). For {\bf b)} and {\bf c)}, the TS is broken spontaneously, namely, broken by the ground state: the latter is not supersymmetric, as follows from the fact that its eigenvalue is non-zero. A non-supersymmetric ground state is easily ``excitable'' because it is doubly degenerate. This is the precursor of the Goldstone theorem in spatially extended models, ensuring the presence of gapless excitations. 
		%responsible for emergent long-range correlations related to the 1/f noise. 
		The corresponding spontaneous order is the embodiment of the stochastic generalization of the butterfly effect.
	}
\end{figure}

After this brief but important prelude, we are now in a position to demonstrate that chaos is the spontaneous breaking of TS. First, let's introduce the concept of the ground state of the operator $\hat H$. This is one of the states with the minimal real part of its eigenvalue. 
%Such state grows faster than others in the limit of infinitely long evolution. 
It represents what is often called ``sustained dynamics'' -- the state of the system that has been allowed to evolve for a very long time without perturbations. From the dynamical systems theory point of view, the ground state is the stochastic generalization of the concept of an attractor. From the Quantum Field Theory perspective, the ground state (and the lowest excitations above it) defines the effective description of the system in the long-wavelength limit known as the effective field theory, which in the case of STS is the theory of the butterfly effect.\cite{OVCHINNIKOV2024114611}

Second, let's recall that there are only three types of spectra of a pseudo-Hermitian operator such as $\hat H$, as shown in Figs.~\ref{Figure_3}a-c.\cite{Mos023} In the case of Fig.~\ref{Figure_3}a, the ground state is one of the zero-eigenvalue supersymmetric singlets. By definition, when the ground state is symmetric the corresponding symmetry is unbroken. On the other hand, for the spectra in Figs.~\ref{Figure_3}b and c, the ground state eigenvalue is nonzero, which means it is not supersymmetric. In other words, the symmetry of the equations of motion is no longer shared by the ground state: SSB has occurred. However, we must stress here that the concept of SSB -- the situation when the ground state has a lower symmetry than that of the evolution operator -- should not be confused with the \emph{explicit} symmetry breaking. In the latter case, the symmetry of the evolution operator is broken by an external perturbation. In the context of TS, this situation may occur in discrete-time dynamics where the corresponding maps are not smooth and the corresponding transformations of wavefunctions may not commute with the exterior derivative. In Ref.\cite{Max_2019}, it was suggested that it is the explicit symmetry breaking of TS that allows chaos to occur in discrete-time maps of dimensionality lower than three.

Note also that, for the spectra in Figs.~\ref{Figure_3}b and c, the real part of the ground state eigenvalue is negative. This has dramatic consequences: the number of periodic solutions grows exponentially with time, which is a fundamental characteristic of chaotic dynamics.\cite{Gilmore_book} This can be easily seen from the dynamical partition function $Z_{tt'}$, which is the trace of the finite-time evolution operator $\hat {\mathcal M}_{tt'}$. %This dynamical partition function counts the number of periodic trajectories in the long-time limit. 
From Eq.~(\ref{GTO}) it is easy to see that such a quantity grows exponentially at long times,
\begin{eqnarray}
	\left.Z_{tt'}\right|_{t-t'\to\infty}  = \left.\mathrm{Tr} \hat {\mathcal M}_{tt'}\right|_{t-t'\to\infty} \propto e^{ (t-t')\Delta },\;\;  \Delta = -\mathrm{Re}H_{0}. \label{StochasticChaos}
\end{eqnarray}
where $H_{0}$ is the eigenvalue of the ground state, and the rate of growth $\Delta$ is known in dynamical systems theory as ``pressure''.\cite{Rue02} If the pressure is zero, which corresponds to unbroken TS in Fig. \ref{Figure_3}a, then there is no exponential growth of the partition function and, consequently, of the periodic solutions. If $\Delta >0$, which is the case of spontaneously broken TS in Figs. \ref{Figure_3}b and c, there is such an exponential growth. 
%One object that can tell, for instance, chaotic vs. non chaotic dynamics are the partition function, i.e., the trace of GTO. It has the meaning of the number of periodic solutions (averaged over noise configurations). From the dynamical systems point of view, the key feature of chaotic dynamics are the exponential growth of the number of (unstable) periodic solutions in the limit of long evolution. {\bf REF Gilmore's book} Accordingly, chaos in stochastic models can be defined as a situation when the spectral radius of GTO is larger then unity or, equivalently, when the lowest real part of eigenvalues is negative\cite{Rue02}
%\begin{eqnarray}
	%    \left.Z_{tt'}\right|_{t-t'\to\infty}  = \left.\mathrm{Tr} \hat {\mathcal M}_{tt'}\right|_{t-t'\to\infty} \propto e^{ (t-t') \Delta}, \Delta = - \min_i  Re (H_i), \label{StochasticChaos}
	%\end{eqnarray}
%where $\Delta$ is known as pressure and positive pressure signifies chaotic dynamics. 
In other words, SSB is equivalent to stochastic chaos defined as positive pressure. 

In fact, positive pressure is only one of the possible definitions of stochastic chaos in dynamical systems theory, and its equivalence with SSB makes it the most natural from the physical point of view. Moreover, since TS is really a mathematical concept from Algebraic Topology, positive pressure and/or SSB are the most natural definition of stochastic chaos from a mathematical point of view as well. In addition, there are other advantages of this definition of stochastic chaos, \cite{Max_2019} including the stochastic generalization of the Poincar\'e-Bendixson theorem \cite{OvcEntropy} which states that only models with $D>2$ can be chaotic.\cite{Gilmore_book}

\subsection{The Phase Diagram of Chaotic Dynamics} 
%%%%%%%%%%%%%%%%%%%%%%%%%%%%%%%%%%%%%%%%%%%%%%%%%%%%
%%%%%%%%%%%%%%%%%%%%%%%%%%%%%%%%%%%%%%%%%%%%%%%%%%%%
%One of the low hanging fruits from STS -- the basic phase diagram addressed here -- comes from analysis of the internal structure of the global ground state. Most certainly, this topic will have many details in the future. At this moment, we would like to briefly discuss what is already understood in this direction. 
%Finally, the picture of chaos in terms of SSB offers a complete classification of chaotic dynamics in stochastic models (see Fig. \ref{Figure_2}). This can be explained as the phase diagram of the dynamics in terms of both the noise strength and other parameters of a physical system. 
\begin{figure}[t]
	\begin{center}
		\noindent\includegraphics[width=1\linewidth]{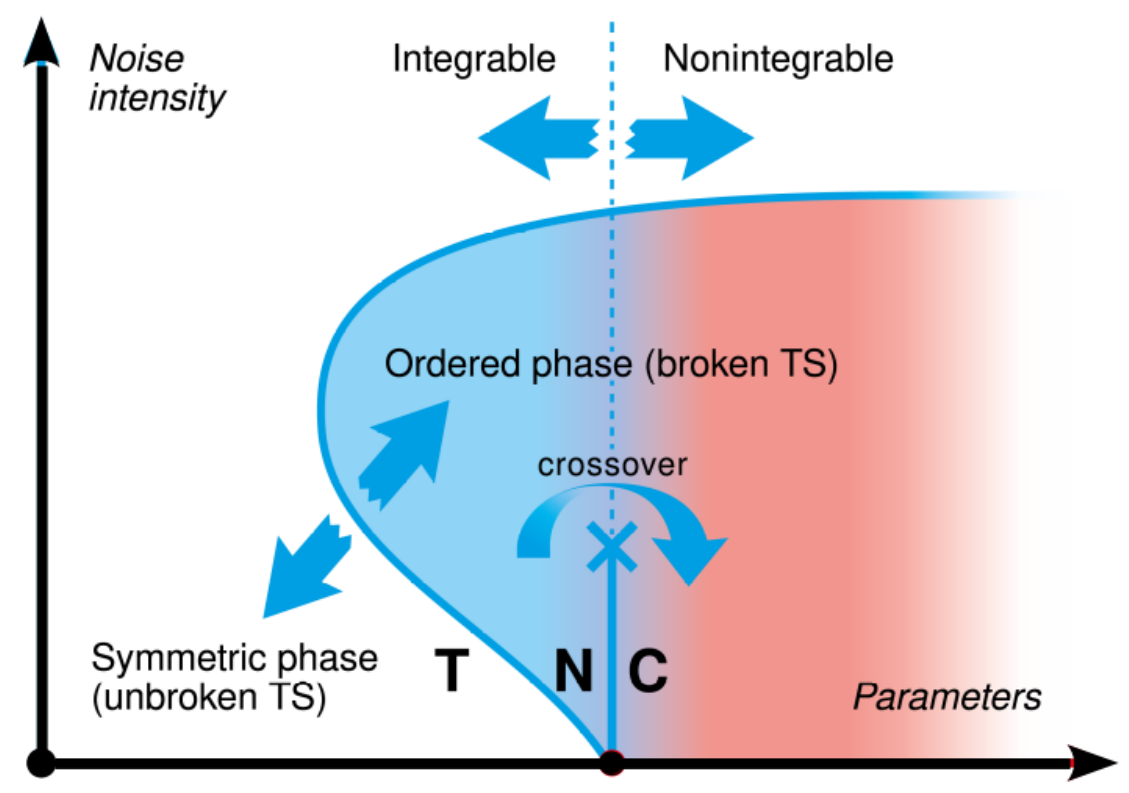}
	\end{center}
	\caption{\label{Figure_4} {\bf Phase diagram of chaotic dynamics} The basic phase diagram of stochastic dynamics in terms of the noise intensity and other parameters of the model (vertical and horizontal axes, respectively) with respect to whether the topological supersymmetry (TS) is spontaneously broken or unbroken, and whether the flow vector field is integrable or non-integrable/chaotic. The symmetric phase with unbroken TS is denoted as (T). At low noise, the ordered non-integrable phase can be referred to as chaotic (C), as it hosts conventional deterministic chaos. The ordered integrable phase can be identified as the noise-induced chaos (N) or the ``edge of chaos''. The dynamics in the N-phase is dominated by noise-induced instantons/anti-instantons and in the deterministic limit it collapses onto the border of the conventional deterministic chaos. At stronger noise, the transition between C and N phase turns into a crossover as the noise smears out the boundary between different instantons. As the noise intensity increases even further, TS must eventually be restored irrespective of the integrability of the flow.}
\end{figure}

Finally, the picture of chaos in terms of SSB allows us to identify the general properties of the phase diagram of chaotic dynamics in terms of both the noise strength and other parameters of the system (see Fig. \ref{Figure_4}). 

The classification of deterministic models in regard to TS breaking is very simple. A deterministic model is either chaotic or not depending on its flow vector field $F$ in Eq.~(\ref{SDE}).\cite{Gilmore_book} In dynamical systems theory, a flow is called integrable if its global unstable manifolds of all dimensionalities (including all attractors) are well-defined topological (sub-)manifolds of the phase space.\cite{Gilmore_book} This is in contrast with flows that are chaotic or non-integrable. We can then say that in deterministic models TS is broken by the non-integrability of the flow vector field, and, particularly, its non-supersymmetric ground state represents an attractor which is not a topological manifold from the point of view of algebraic topology  -- a ``strange attractor'' (see Figs.~\ref{Figure_1} and~\ref{Figure_2}).\cite{Gilmore_book}

In the presence of (weak) noise, however, TS breaking and the loss of integrability do not happen simultaneously (Fig. \ref{Figure_4}). Particularly, TS breaking may precede the non-integrability of the flow giving rise to a new noise-induced phase where TS is spontaneously broken by a mechanism different from non-integrability. This mechanism is the most reliable mechanism of supersymmetry breaking in high-energy physics.\cite{DynSusyBrWitten} Namely, it is the condensation of instantons\cite{Solitions_Review_1} or, more accurately, instantons matched by their time-reversal noise-induced counterparts called anti-instantons.

Instantons -- the ``quanta'' of the nonlinear dynamics\cite{TFT_BOOK,Solitions_Review_1,Polyakovinst} -- are dynamical processes comprising trajectories that connect distinct critical points (or manifolds more generally) of the flow that can be loosely recognized as states of (un)stable equilibria. Real life examples of (composite) instantons include balloon popping, neuronal avalanches, protein folding, impact fragmentation, and many others including the life cycle of a living organism. Noise-induced instantons dominate dynamics in this new phase which can be called {\it noise-induced chaos}. This phase collapses onto the critical border of conventional chaos in the deterministic limit. Therefore, this theoretical framework helps us understand the phenomenon known in the literature as the ``edge of chaos.''\cite{Packard1988,DynamicalComplexity,Crutchfield} 

In fact, it was speculated that this region of noise-induced chaos could be employed for combinatorial optimization, in view of both its dynamical long-range order (DLRO) and integrability of the flow vector field. \cite{Optimization} This is in line with previous suggestions that the edge 
of chaos could be ideal for both processing and storing of information.~\cite{edge-chaos} On the other hand, the TS is effectively broken on instantons, giving rise to DLRO, while also maintaining integrability of the flow.~\cite{DMM_1,DMM_2} This type of DLRO is exploited in the new computing paradigm of memcomputing which has been extensively used to solve a wide variety of combinatorial optimization problems.~\cite{di2022memcomputing} However, the effective breakdown of TS on instantons is distinct from   
its spontaneous breakdown we have discussed all along. In other words, ``computing with instantons'' is not the same 
as computing at the ``edge of chaos''.

Finally, the picture of the basic phase diagram is completed by pointing out that in the limit of strong noise TS is eventually restored, just like any other symmetry-breaking orders --such as crystal structure, ferromagnetism, etc.-- disappear at high temperatures (Fig. \ref{Figure_4}).

%It is clear that for spectra b and c, the ground states have non-zero eigenvalues and, consequently, are non-supersymmetric. By definition, this imply that TS is broken spontaneously, that is, by the ground state. In other words, stochastic chaos of the dynamical systems theory is equivalent to the spontaneously broken TS of STS. This is not just a formal equivalence. For instance, spontaneously broken TS is the link to the Goldstone theorem: when TS is spontaneously broken, the ground state is degenerate and the system can be effortlessly excited. In higher-dimensional theories, this degeneracy evolves into a gapless branch of excitations above the ground state called goldstinos. This qualitatively explains ubiquitous phenomenon called 1/f noise, as we will shortly dicuss in Sec.\ref{Sec1fnoise} below. 

%%%%%%%%%%%%%%%%%%%%%%%%%%%%%%%%%%%%%%%%%%%%%%%%%%%%%%%%%%%%%%%%%5
%%%%%%%%%%%%%%%%%%%%%%%%%%%%%%%%%%%%%%%%%%%%%%%%%%%%%%%%%%%%%%%%%5
\section{A Supersymmetry-Broken World}
%%%%%%%%%%%%%%%%%%%%%%%%%%%%%%%%%%%%%%%%%%%%%%%%%%%%%%%%%%%%%%%%%5
%%%%%%%%%%%%%%%%%%%%%%%%%%%%%%%%%%%%%%%%%%%%%%%%%%%%%%%%%%%%%%%%%5

%%%%%%%%%%%%%%%%%%%%%%%%%%%%%%%%%%%%%%%%%%%%%%%%%%%%%%%%%%%%%%%%%5
%%%%%%%%%%%%%%%%%%%%%%%%%%%%%%%%%%%%%%%%%%%%%%%%%%%%%%%%%%%%%%%%%5
%\subsection{Philosophy/General}
%%%%%%%%%%%%%%%%%%%%%%%%%%%%%%%%%%%%%%%%%%%%%%%%%%%%%%%%%%%%%%%%%5
%%%%%%%%%%%%%%%%%%%%%%%%%%%%%%%%%%%%%%%%%%%%%%%%%%%%%%%%%%%%%%%%%5
Recognizing chaos as a spontaneous order due to supersymmetry breaking is not just a substantial change of perspective. It has also deep implications for our understanding of other phenomena. A useful concept that can be used to see this is effective field theory (EFT), i.e., an approximate theory that describes slow degrees of freedom and accounts for the fast degrees of freedom effectively. It is closely related to the already introduced concept of ground state representing the ``sustained dynamics'': the EFT describes these dynamics and low-lying excitations above them.

As a consequence of the Goldstone theorem, the EFT of (spatially-extended, translationally-invariant) chaotic models is a theory of gapless ``goldstinos,'' the fermionic analogue of the low-frequency Nambu-Goldstone excitations that emerge in spontaneously broken symmetries.\cite{Book_Peskin} Due to the gaplessness of goldstinos, the EFT must be scale-invariant,\cite{RFT_SSB_book} or, a conformal field theory (CFT) with some correlators being long ranged.\cite{Intro-CFT} This qualitatively explains the widespread occurrence of long-range behavior in chaotic dynamics, known as $1/f$ noise.\cite{Keshner_1_f_noise_1982,Voss_1979,RevModPhys.53.497}

Moreover, the goldstinos represent the differentials of the ``wavefunctions'' understood as differential forms, Eq.~(\ref{psi}). As we have discussed, the fermions are the objects that determine the Lyapunov exponents\cite{Lyapunov_SUSY} which, in turn, characterize the butterfly effect. In other words, the EFT is the theory of the butterfly effect.\cite{OVCHINNIKOV2024114611} The EFT is scale-invariant and in some cases there may even exist a possibility for a holographic description of the butterfly effect.\cite{OVCHINNIKOV2024114611} 
%Without getting into details, it must be pointed out that such an EFT of the butterfly effect would be a first consistent physical theory of the BE.  
Since the butterfly effect is essentially a long-range memory, such EFT may help identify/uncover long-range information in various realizations of chaotic dynamics in the Universe. 

In fact, the Universe as a whole can be viewed as a gigantic dynamical system harboring a ``hidden'' global long-range order, as suggested by countless scale-free behaviors\cite{Asc11,Zong_Kuan_Guo_2011} pointing out that the corresponding TS is broken -- either spontaneously or because cosmic evolution can be interpreted as an enormous instanton that began at the Big Bang. Although our present understanding of this picture remains speculative, we believe it offers a conceptual direction worthy of further discussion. 

%%%%%%%%%%%%%%%%%%%%%%%%%%%%%%%%%%%%%%%%%%%%%%%%%%%%%%%%%%%%%%%%%5
%%%%%%%%%%%%%%%%%%%%%%%%%%%%%%%%%%%%%%%%%%%%%%%%%%%%%%%%%%%%%%%%%5
\section{Conclusion}
%%%%%%%%%%%%%%%%%%%%%%%%%%%%%%%%%%%%%%%%%%%%%%%%%%%%%%%%%%%%%%%%%5
%%%%%%%%%%%%%%%%%%%%%%%%%%%%%%%%%%%%%%%%%%%%%%%%%%%%%%%%%%%%%%%%%5

In conclusion, we have provided a short account of the phenomenon of chaos seen through the lens of the supersymmetric theory of stochastics (STS). This unusual, yet rigorous point of view reveals that dynamical chaos represents an {\it ordered} phase of dynamical systems, associated with the spontaneous breakdown of topological supersymmetry, a symmetry that is inherently present in all natural dynamical systems. In fact, since chaos may not be the best identifier for any order, the word {\it chronotaxis} was suggested in Ref.~\cite{Max_2019} as a better representation of this phenomenon. 

This surprising reinterpretation of chaos offers then a compelling and elegant explanation of phenomena that still generate a lot of discussion. For instance, the ubiquitous $1/f$ noise emerges as a direct consequence of the Goldstone theorem, while the ``edge of chaos'' is understood as a special form of chaotic behavior on the border of conventional chaos in which topological supersymmetry is broken by instantonic processes.\cite{OvcEntropy} % In view of the fact that topological supersymmetry is common to all (stochastic) differential equations, we then expect that many other phenomena which showcase dynamical long-range order may originate from the spontaneous breaking of such a symmetry.   
%Furthermore, STS has already found some practical applications in unexpected fields such as neurodynamics, and the unconventional computing paradigm known as memcomputing, where it has revealed topological aspects of neuronal activity and computational efficiency, respectively, otherwise difficult to infer from traditional approaches. 
Beyond these conceptual insights, the STS framework has already found its usefulness in several  applications.\cite{Torsten,li2018,di2022memcomputing} 

%astrophysics,\cite{Torsten} neurodynamics,\cite{li2018} and unconventional computing paradigms.\cite{di2022memcomputing}
Given the universal applicability of stochastic dynamics, it can be anticipated that the relevance of this new picture of chaos will continue to expand across diverse scientific domains. A particularly important step is the development of a systematic methodology for constructing effective field theories of chaos,\cite{OVCHINNIKOV2024114611} that would provide a consistent theoretical description for the butterfly effect, in a manner analogous to, e.g., how the Ginzburg–Landau theory formalized superconductivity.~\cite{Tinkham2004}

There is yet another interesting aspect of the TS-breaking picture of chaotic dynamics. As we pointed out, the ground states of chaotic systems are not just probability distributions but they have nontrivial fermionic content, whose low-lying excitations  describe the butterfly effect. This situation is somewhat similar to superconductors, where the low-lying excitations are related to the phase of the order parameter and not its amplitude, which is related to the probability distribution.~\cite{Tinkham2004} In view of this analogy, can we then expect chaotic dynamics to exhibit analogous phenomena as the Josephson effect, Berry phase, or even entanglement? We hope that, by building on this new understanding of chaos, future research will shed more light on these interesting questions.

On a more speculative level, this new viewpoint on chaos is particularly intriguing in the context of neurodynamics. In fact, the observed avalanche-dominated brain dynamics~\cite{beggs2003} may be a manifestation of noise-induced chaos. If that is the case, the brain dynamics may have a hidden long-range topological order~\cite{Jay} that may play an important role in short-term memory and, consequently, consciousness.

Finally, as we have briefly outlined in this perspective, STS shares many similarities with Quantum Theory. Therefore, besides its practical applications, it may be useful from a broader perspective by bridging the divide that typically separates communities of researchers working on the theory of classical complex systems and other areas of theoretical physics, from quantum field theory to string theory. We therefore hope this perspective will spark the reader's curiosity and promote dialogue and collaboration across traditionally isolated disciplines. 

%may pave the way for constructing rigorous, high-level mathematical frameworks for other complex fields -- such as neurodynamics -- potentially elevating them to the same theoretical depth as condensed matter physics.
%Beyond these specific results, STS may have a broader and subtler -- but no less important -- impact on the structure of science itself. Today, mathematical physics is largely split into two domains: quantum and classical. The gap between them is substantial, often making collaboration between, for example, dynamical systems theorists and string theorists difficult. STS offers a unified mathematical language that could help bridge this divide, 

\section*{Acknowledgments}
%%%  Use this section to acknowledge contributions 
%%%  from non-authors and list funding sources, 
%%%  including grant numbers.
We would like to thank Kang L. Wang, Savdeep S. Sethi, Gabriel A. Weiderpass, Ben Isreali, Daniel Toker, Cheng-Zong Bai, Eugene Ingerman, and Fabio L. Traversa, who have positively influenced this work, and Anna Afinogenova for assistance with the preparation of the figures. M.D. acknowledges support from the National Science Foundation via grant No. ECCS-2229880, and the Alexander von Humboldt Stiftung through the Humboldt Research Award. 

\bibliography{references}
\end{document}